\documentclass[prb,aps,amsmath,showpacs,preprint,]{revtex4-1}
\usepackage{graphicx}
\usepackage{epsfig}
\usepackage{multirow}
\usepackage{dcolumn}

\usepackage{bm}
\usepackage{subfigure}
\usepackage{color}

\begin{document}

\title{Thermoelectric properties of chalcopyrite type CuGaTe$_2$ and chalcostibite CuSbS$_2$}

\author{Vijay Kumar Gudelli and V. Kanchana$^{*}$}
\affiliation {Department of Physics, Indian Institute of Technology Hyderabad, Ordnance Factory Estate, Yeddumailaram-502 205, Andhra Pradesh, India}
\author {G. Vaitheeswaran}
\affiliation {Advanced Centre of Research in High Energy Materials (ACRHEM), University of Hyderabad, Prof. C. R. Rao Road, Gachibowli, Hyderabad-500 046, Andhra Pradesh, India}
\author {A. Svane and N. E. Christensen}
\affiliation {Department of Physics and Astronomy, Aarhus University, DK-8000 Aarhus C, Denmark} 
\date{\today}

\begin{abstract}

Electronic and transport properties of CuGaTe$_2$, a hole-doped ternary copper based chalcopyrite type semiconductor, 
are studied using calculations within the Density Functional Theory and solving the Boltzmann transport equation within the 
constant relaxation time approximation. The electronic bandstructures are calculated by means of the full-potential linear augmented plane wave method, 
using the Tran-Blaha modified Becke-Johnson potential.  The calculated band gap of 1.23 eV is in agreement
 with the experimental value of 1.2 eV. The carrier concentration- and temperature dependent thermoelectric properties of CuGaTe$_2$ are derived, and a
 figure of merit of $zT=  1.69$ is obtained at 950 K for a  hole concentration of $3.7\cdot10^{19}$ cm$^{-3}$, in agreement with a recent experimental finding of $zT=  1.4$, confirming that CuGaTe$_2$ is a promising material for high temperature thermoelectric applications. The good thermoelectric performance of p-type CuGaTe$_2$ is associated with anisotropic transport from a combination of heavy and light bands. 
Also for CuSbS$_2$ (chalcostibite) a better performance is obtained for p-type than for n-type doping.  
The variation of the thermopower as a function of temperature and concentration suggests that CuSbS$_2$ 
will be a good thermoelectric material at low temperatures, similarly to the isostructural CuBiS$_2$ compound.
 
\end{abstract}

\maketitle


\section{Introduction}
	Thermoelectric (TE) materials with potential applications  within power generation and refrigeration have represented a thrust area of research 
	for the past few decades. 
	TE materials can convert waste heat into electric power and hence 
play a vital role in meeting the present condition of energy crisis and environment pollution.\cite{Majumdar,Bell,Snyder,Dress} 
The performance of a TE material is reflected in the
dimensionless figure of merit, $zT$, given by $zT = \frac{S^2\sigma T}{\kappa}$, where $S$, $\sigma$, $\kappa$ and $T$ are the thermopower, 
the electrical conductivity, the thermal conductivity, and the absolute temperature, respectively. 
$\kappa$ includes both the electronic, $\kappa_e$, and the lattice contributions,
$\kappa_l$, i.e. $\kappa=\kappa_e+\kappa_l$. From this  expression, it is evident that finding materials with high $zT$ is a  challenge, 
as it appears that such a material should  satisfy the conflicting requirements of high thermopower, 
which is often found for doped insulators, and behave as a
good electrical conductor like a metal with a low thermal conductivity. Good electrical conductivity 
and poor thermal conductivity implies a weak electron scattering and strong phonon scattering. 
Remarkable progress has been made in recent years exploring different classes of materials for better TE 
performance.\cite{Snyder, Dress, Georg, singh, Abdeljalil, Jiong, andrew, Wang, Jovovic, David, YuLi, khu, Lijun, Parker, Khuong, Mun} 
Few materials with $zT > 1$ are known.\cite{Snyder} Apart from the material
properties necessary for a good figure of merit, there are also other materials properties to be considered. 
The materials should possess a high melting point as far as $"$waste heat recovery$"$ is concerned,
 and they should be structurally stable in the operating temperature range. 
In addition, the constituents of the materials should be abundant and in-expensive. 
The real success in the field of thermoelectrics lies in identifying a material with all the desired properties. There are few
commonly used good TE materials such as PbTe and Bi$_2$Te$_3$.\cite{Satterwaithe, Goldsmid, DJS}  
Bi$_2$Te$_3$ can be hole doped and electron doped and has $ zT \sim 1$ at room temperature, but still it  has a shortfall as Te is a rare element and has
many restrictions for large-scale applications.\cite{cubis2, Amatya} Recently, alternative Pb-free materials such as AgGaTe$_2$ and CuBiS$_2$, 
have been considered.\cite{cubis2, Wentao, aggate2} The present work on the chalcopyrite semiconductor CuGaTe$_2$ is motivated by
the recent experimental work \cite{ADV. MATER}  reporting a $zT$ of 1.4 at temperatures above 800 K for this compound. 
Although many of the PbTe based materials may also have  $zT > 1.5$, their range of operating temperature is much lower
than that of CuGaTe$_2$.\cite{K. F. Hsu,Y. Pei,Biswas,Poudeu,Androulakis,Pei} Ab-inito calculations for the similar chalcopyrite compound AgGaTe$_2$ 
were performed by  Parker and Singh\cite{cubis2}  and  Wu et al.,\cite{Wentao} who showed that hole doping of
AgGaTe$_2$  improves its TE performance. The present work is a theoretical analysis of CuGaTe$_2$, 
which supports the classification of this compound as an excellent thermoelectric material. 
 In addition,  also  CuSbS$_2$ is examined. This compound is isostructural with CuBiS$_2$, which has been predicted to be an excellent TE material.\cite{cubis2}

 The paper is organized as follows: Section II  describes  the method used for the calculations, and section III presents the results and
a discussion. Conclusions are given in section IV.    

\section{Methodology}

The electronic band structures were calculated by means of the full-potential linear augmented plane wave (FP-LAPW) method based on 
first-principles density functional theory as implemented in the WIEN2k code.\cite{Blaha}  
Since calculations using standard local-density (LDA) or
Generalized Gradient Approximation (GGA) schemes for the exchange-correlation potential underestimate the band gaps of semiconductors, we have used the 
modified GGA known as the Tran-Blaha modified Becke-Johnson\cite{Becke} potential 
(TB-mBJ).\cite{Tran1}  For k-space integrations a 20x20x20 k-mesh was used
for CuGaTe$_2$ and 19x31x8 k-mesh for CuSbS$_2$ in the Monkhorst-Pack scheme, resulting in 641 and 800 k-points in the irreducible parts 
of the Brillouin zones for the two compounds, respectively. The self-consistent calculations included  spin-orbit coupling. The crystal
structure of CuGaTe$_2$ is tetragonal with space group $I\bar{4}2d$  (no. 122) and lattice parameters a=6.028 \AA\ and c=11.949 \AA.\cite{Kuhn} For CuSbS$_2$, 
the crystal structure is in the orthorhombic space group $Pnma$ (no. 62) with lattice parameters a=6.018 \AA, b=3.7958 \AA\  and c=14.495
\AA.\cite{Atsushi}
 All the calculations were performed with the experimental lattice parameters with an energy convergence criterion of $10^{-6}$ Ry per formula unit.  
 The carrier concentration (p for holes and n for electrons) and temperature ($T$) dependent thermoelectric properties like thermopower ($S$), electrical
conductivity ($\sigma$), power factor ($S^2$$\sigma$), and figure of merit ($zT$) were calculated using the BOLTZTRAP\cite{Madsen} code, within the 
Rigid Band Approximation (RBA)\cite{Scheidemantel,Jodin} and the constant scattering time ($\tau$) approximation (CSTA). In the RBA
  the band structure is assumed unaffected by doping, which only leads to a shift of the chemical potential. For semiconductors it is a good approximation 
 for calculation of the transport properties, when the doping level is not too high.\cite{Jodin,Chaput,Bilc,Ziman,Nag,Mazin}
In the CSTA, the scattering time of electrons is assumed independent of the electron energy, while it may depend on carrier concentration and temperature. 
A detailed discussion of the CSTA is given in Refs. \onlinecite{singh}, \onlinecite{aggate2} and  \onlinecite{Khuong}, and references therein. The only situation
where the CSTA can fail is when bipolar conduction is significant, which happens in narrow-gap materials. 
According to Sofo and Mahan,\cite{sofo} the best performance of a 
thermoelectric material is found when the energy gap is about 10 k$_B$T$_o$, where T$_o$ is the operating temperature, 
as far as direct band gap materials are concerned. In the case of
CuGaTe$_2$, which is found to have a direct band gap of about 1.2 eV, this corresponds to T$_o$ $\sim $1400 K. 
According to the present calculations $zT$ reaches its maximum value near 950 K, and this justifies the approximation  used in the 
calculations for CuGaTe$_2$, as we are still far away from the region of bipolar conduction. In the case of CuSbS$_2$, 
the band gap around 1 eV corresponds to T$_o$ $\sim$ 1200 K, however the melting point is\cite{wachtel} only 825 K. 
Therefore, the present calculations for CuSbS$_2$ cover only temperatures up to 700 K,  again  a
safe regime  as far as the CSTA is concerned.   

\section { Results and Discussion } 


\subsection{Band structure and Density of States of CuGaTe$_2$ and CuSbS$_2$}
The calculated band structure of CuGaTe$_2$ along the high symmetry directions of the tetragonal Brillouin zone is shown in FIG. 1(a).  
The valence band maximum (VBM) (zero energy in the band structure plot) and the conduction band minimum (CBM) are
both located at the centre of
the Brillouin zone i.e. at the $\Gamma$-point, making the compound a direct-band-gap semiconductor. 
The band gap values obtained within the GGA and the TB-mBJ are given in Table I. From the large band gaps it
is expected that there should not be any bipolar conduction.  
The band structure in the vicinity of the VBM exhibits a mixture of heavy and light bands, which is often favourable for thermoelectric performance, 
one can also see here such a situation in the vicinity of the  VBM. 
The heavy band found just below the VBM  arises from the Cu-d and Te-p states, 
and below this lies the light band of Te-p and Ga-d character. The Density of States (DOS) of CuGaTe$_2$ is
shown in FIG. 1(b), where it is apparent that the major contribution to the bands at the VBM comes from the Cu-d states, 
leading to a strong increase in the DOS as energy is moving away from the VBM. 
Similarly, the DOS also rises steeply above the CBM, albeit not as distinctly as around the VBM, cf. 
 FIG. 1(a) and 1(b). 
The heavy bands usually contribute to a high thermopower while the lighter bands offer an
advantage of high mobility, a favorable combination often leading to an excellent TE performance of the material. 
Although one might also expect that n-doping could give good thermoelectric performance for CuGaTe$_2$, 
the presence of multiple heavy bands at the VBM would likely favour the p-type doping over n-type, 
as is indeed found in the succeeding subsection, where   the transport properties and
 optimized doping level  are discussed.

 For  CuSbS$_2$ the calculated band structure along the high symmetry directions of the orthorhombic Brillouin zone is shown in FIG. 2(a).  
 This compound is  an indirect-band-gap semiconductor with a band gap of  1.05 eV, where experimental gap values are somewhat higher, 
 1.38 eV\cite{Zhou} and 1.52 eV\cite{Rodriguez} (see Table I).
 The gap is large  enough to prevent bipolar
conduction for operating temperatures below the melting point.  
The bands near the $\Gamma$-point are dispersive in all three symmetry directions, albeit least dispersive along $\Gamma$-Z. 
The 
 DOS of CuSbS$_2$ is shown in FIG. 2(b), from which it is seen that the major contribution to the DOS near the VBM comes from the Cu-d states. 
 Compared to the DOS of CuGaTe$_2$, there is less symmetry between the CBM and  VBM regimes, which would render p-type doping more favorable over n-type doping  in
CuSbS$_2$ than in CuGaTe$_2$.
The presence of the heavy mass band at the VBM for both compounds indicates the possibility for  excellent thermo-electric performance.  This is discussed in the following section.

\subsection{Thermoelectric properties of CuGaTe$_2$ and CuSbS$_2$}
 The carrier concentration and temperature dependent thermoelectric properties of CuGaTe$_2$ are obtained by solving the 
 Boltzmann transport  equation as implemented in the  BOLZTRAP\cite{Madsen} code. The calculated
properties are the thermopower, the electrical conductivity divided by the 
scattering time (i.e. $(\sigma$/$\tau)$), the power factor, and the figure of merit as functions of carrier concentration and temperature.
Since most of the experiments are done in poly-crystalline samples, we have calculated the averages of the thermopower and the electrical conductivity (respectively) over three orthogonal axes in order to estimate the figure of merit. 
Due to the lack of experimental data for CuSbS$_2$ from which the relaxation time may be extracted (temperature dependent concentration and resistivity), 
we have for this compound only studied  the concentration and temperature dependent thermopower, and concentration dependent $(\sigma$/$\tau)$ ratio. 


\subsubsection{\bf Thermopower}

The calculated value of $S$ depends on carrier
concentration and temperature, but it is independent of $\tau$  due to the CSTA. 
The calculated thermopower for CuGaTe$_2$, $S(T,p)$,  as a function of hole concentration $p$ at different temperatures along the a- and c-axes is shown in Figs. 3(a) and 3(b). 
The trend of the thermopower 
 along the two axes is similar to what has been found for other thermoelectric materials with the tetragonal structure.\cite{Khuong, aggate2}
 It is also seen that the thermopower increases with decreasing carrier concentration.
 The Pisarenko behavior, i.e. logarithmic variation of the thermopower with carrier concentration, is found in the range of $p=10^{18}$--$10^{20}$ cm$^{-3}$, 
which is an optimum working region for good thermoelectric materials. 
The variation of the thermopower of CuGaTe$_2$ as a function of temperature was measured by Plirdpring {\it et al.}\cite{ADV. MATER} These authors also determined the temperature dependent
carrier density $p(T)$, so by combining this information with the calculated $S(T,p)$, the calculated temperature dependent thermopower $S(T)\equiv S(T,p(T))$ may be obtained
and compared directly to experiment. This is
done in FIG. 4.  
The theory and experiment are in good agreement with a qualitatively similar shape and a maximum thermopower around 400 K (maximal thermopower of 450 $\mu$V/K and 410 $\mu$V/K in 
theory and experiment, respectively). At 900 K the thermopower has fallen to around 250 $\mu$V/K. At 950 K, where the measured figure of merit has its maximum value 
(about $zT = 1.4$), an experimental thermopower of $S=244 \mu$V/K
is measured,\cite{ADV. MATER} as compared to the theoretical value of $S=287 \mu$V/K. 

  
It is  interesting to examine the thermopower also in the case of electron doping, which is illustrated in FIG. 5, showing the thermopower variation with electron concentration at different
temperatures and along both the a and c axes. The magnitude of the thermopower is about the same for given carrier density for p- and n-type doping, about 400 $\mu$V/K at
$10^{19}$ cm$^{-3}$ and 900 K, however the anisotropy is larger in the case of n-type doping. From FIG. 5 the difference in the thermopower
along $a$ and $c$ is $\sim$ 65 $\mu$ V/K (larger along $a$), which is about twice as large as in p-type doped  CuGaTe$_2$ (FIG. 3). 
This might be unfavorable for the potential use of n-type doped CuGaTe$_2$ as a thermoelectric material.\cite{aggate2}
   The variation of the thermopower with temperature for CuSbS$_2$ at selected  hole concentrations is displayed in FIG. 6(a). 
   The thermopower increases with decreasing carrier concentration as in all good TE materials. 
The melting point of CuSbS$_2$ is relatively low, $\sim 825$ K,\cite{wachtel} and hence this compound is less suited for high 
temperature applications, however it may find  application as a cooling component, similar to the isostructural CuBiS$_2$ compound.\cite{cubis2}  
To investigate the anisotropy of thermoelectric properties we have calculated the directional dependent thermopower as function of the
hole and electron concentration as shown in FIG. 6(b, c). From the figure, it appears that the thermopower is very similar in the   
the $x$ and $z$ directions, while it is higher in the $y$ direction by about 70 $\mu$V/K  for hole doping and numerically lower by 100 $\mu$V/K for electron doping. 
Thus, the thermopower for hole doping is somewhat larger than for electron doping, assuming similar carrier densities. The magnitude of the thermopower for hole doping is similar  to what is found for the isostructural CuBiS$_2$,\cite{cubis2}
while the anisotropy is somewhat larger.

\subsubsection{\bf Electrical conductivity}

The electrical conductivity may be estimated for CuGaTe$_2$ by a combination of theory and experiment. This requires that, the scattering time $\tau$ is 
estimated, which is possible from the experiments of Ref. \onlinecite{ADV. MATER}, if the scattering time is assumed to be a function of temperature alone, i.e. $\tau=\tau(T)$.
The calculated ratio $(\sigma/\tau)$ is a function of $T$ and carrier concentration $p$, i.e. $(\sigma/\tau)(T,p)$. 
In the experiments the carrier concentration itself is a function of temperature and displayed in Fig. 3 of Ref. \onlinecite{ADV. MATER}. 
The experimental data are available up to 800 K, and we have extrapolated in order  to get the concentration up to 950 K.
The scattering time then follows from the relation 
	\[
	\tau(T)=\frac{\sigma(T)_{exp}}{(\sigma/\tau)(T,p(T))}. 
\]
Here, $\sigma(T)_{exp}$ and $p(T)$ are the measured conductivity and carrier concentration, respectively, while $(\sigma/\tau)$ is the calculated ratio. 
Subsequently, we may obtain $\sigma$ as a function of two variables by multiplying the calculated ratio by $\tau(T)$:
	\[
	\label{sigma}
	\sigma(T,p)=\tau(T)\cdot (\sigma/\tau)(T,p).
\]
The scattering time obtained in this way is displayed in figure 7.
A similar procedure was adopted by Ref. \onlinecite {Khuong}, however using only one $p(T)$ value and assuming phonon-dominated scattering. 
Fig. 8  shows the obtained electrical conductivity ($\sigma$)  along the a- and c- directions as  functions of hole concentration.
The electrical conductivity increases  in both cases essentially linearly with the carrier concentration, as also found in Drude's model.
The conductivity is significantly higher along the $a$-axis
than along the $c$-axis, approximately by a factor of 5. 
The electrical conductivity is a decreasing function of temperature for given carrier density, mainly reflecting the temperature dependence of the scattering time.


Having the thermopower and the electrical conductivity, the power factor (S$^2$$\sigma$) is calculated and displayed in FIG. 9 
as function of carrier concentration for three representative temperatures. 
 At 950 K, the calculated power factor at the experimentally realized carrier concentration of $\sim 3.7\cdot 10^{19}$ cm$^{-3}$ (present authors' extrapolation of data from Ref. \onlinecite{ADV. MATER}) is  $ 1.86$ mW/m K$^2$, which is in fair agreement with the corresponding measured value\cite{ADV. MATER} of 1.35 mW/m K$^2$.

In the case of CuSbS$_2$ the calculated variation of the $(\sigma$/$\tau)$ ratio with hole and electron concentration at 300 K  is shown in FIG. 10(a,b). The $(\sigma$/$\tau)$ varies almost 
isotropically with a little difference seen along the b-axis, which is
similar to the trend observed in the thermopower, FIG. 6(b,c). The $(\sigma$/$\tau)$ ratio is higher for hole doping than for electron doping, so for similar carrier density and similar scattering times, this implies better TE performance for p-type CuSbS$_2$. As there are no experimental data available for this compound, we were not able to  proceed  to find the electrical conductivity as in the case of CuGaTe$_2$.

\subsubsection{\bf Thermal Conductivity and $zT$}
      To calculate the figure of merit $zT$ the thermal conductivity $\kappa$ must be known. This was taken from the experiment.\cite{ADV. MATER}
      $\kappa$ is the  sum of the electronic and the lattice thermal conductivities, $\kappa=\kappa_e+\kappa_l$, and these authors showed that if  the electronic
thermal conductivity is obtained from the  Wiedemann-Franz (WF) relation $\kappa_e=L\sigma T$, using the Lorentz
number ($L=2.45\cdot 10^{-8}$ W $\Omega$ / K$^2$), the lattice contribution is well described with an inverse $T$ law, i.e. $\kappa_l=A/T$, 
in the range 400-800 K, as applicable to Umklapp scattering. The WF relation is valid for metals, whereas more accurate modeling may be needed in some 
cases for semiconductors, see Ref. \onlinecite{EES2012} and references therein. At temperatures above 800 K 
the lattice thermal conductivity seems to fall more steeply than dictated by the Umklapp law.\cite{ADV. MATER}
Following Ong et al.\cite{Khuong} (for the case of n-type ZnO) we have found that the experimental 
lattice conductivity is well approximated in the full range 400-950 K by  a fit of the form
 $\kappa= (A-BT)/T$, with $A=2871$ W/m and $B=1.937$ W/m K. 
 This is illustrated in FIG. 11, which also 
 shows the above WF and Umklapp contributions. 



From the experimental information regarding the thermal conductivity, we may 
calculate the figure of merit as function of carrier concentration and temperature as illustrated in FIG. 12. 
FIG. 12(a) shows for three temperatures the carrier dependence of the figure of merit, which is rising above $zT=2$ for carrier densities above $p=10^{20}$ cm$^{-3}$.
Such high carrier densities are not encountered in experiment, so for a better comparison with experiment the $zT$ factor evaluated for the experimental carrier density
(Ref. \onlinecite{ADV. MATER}) is displayed in FIG. 12(b). Here the theoretical figure of merit reaches a value of $zT = 1.69 $ at $T=950$ K, in perfect agreement    
with the  experimental value of $zT = 1.4$ at this temperature.\cite{ADV. MATER} The  
 experimental carrier density at $T=950$ K is only $3.7\cdot 10^{19}$ cm$^{-3}$, as extrapolated from Fig. 3 of Ref. \onlinecite{ADV. MATER}, i. e., significantly lower than the 
optimum density indicated from FIG. 12(a). This high figure of merit indeed renders CuGaTe$_2$ a promising  material for high temperature thermoelectric applications. 
Furthermore, under the same conditions, the Seebeck coefficient is calculated to be 287 $\mu$V/K 
in good agreement with the measured value of 244 $\mu$V/K.\cite{ADV. MATER}

\section{Conclusion}
Electronic transport properties of CuGaTe$_2$ and CuSbS$_2$ were calculated using density functional theory. 
Both  compounds were found to be very good thermoelectric materials with p-type doping. 
A somewhat anisotropic character of the thermopower and electrical conductivity as 
functions of hole concentration was found in both  compounds. 
 In the case of hole doped CuGaTe$_2$ with a concentration of 
$3.7\cdot 10^{19}$ cm$^{-3}$, we obtained a figure of merit $zT=1.69$ and a thermopower of 287 $\mu$V/K at 950 K,
 in excellent agreement with the reported experimental value.
 In the case of CuSbS$_2$ the Seebeck coefficient was found to be large, e.g. at room temperature the calculated 
 thermopower exceeds 400 $\mu$V/K for a hole concentration of $ 10^{19}$ cm$^{-3}$.
 The melting point of this compound is low, and high-temperature thermoelectric applications
 will be impractical, but  
this material might be suitable for low temperature applications. 

\section{Acknowledgement}
V.K.G. and V.K. would like to acknowledge IIT-Hyderabad for providing computational facility. V.K.G. would like to thanks MHRD for the fellowship. V.K. thanks NSFC awarded Research Fellowship for International Young Scientists under Grant No. 11250110051. G.V. thanks Center for Modelling Simulation and Design-University of Hyderabad (CMSD-UoH) for providing access to its computational facilities.

\clearpage
*Author for Correspondence, 
E-mail: kanchana@iith.ac.in

\vspace{2cm}

\begin{table}[h]
\caption{Calculated band gaps within GGA and TB-mBJ, compared to experimental values,  in eV.}
\begin{tabular}{cccc}
\hline
              &      &CuGaTe$_2$     & CuSbS$_2$ \\
\hline

GGA           &      & 0.57         & 0.77       \\

TB-mBJ        &      & 1.23         & 1.05       \\

Expt.         &      & 1.2$^a$      & 1.38$^b$, 1.52$^c$       \\
\hline
\end{tabular}
\end{table}
\begin{center}{$^a$: Ref. \onlinecite{ADV. MATER}}; {$^b$: Ref. \onlinecite{Zhou}}; {$^c$: Ref. \onlinecite{Rodriguez}.}\end{center}

\newpage

\begin{figure}
\begin{center}
\subfigure[]{\includegraphics[width=65mm,height=80mm]{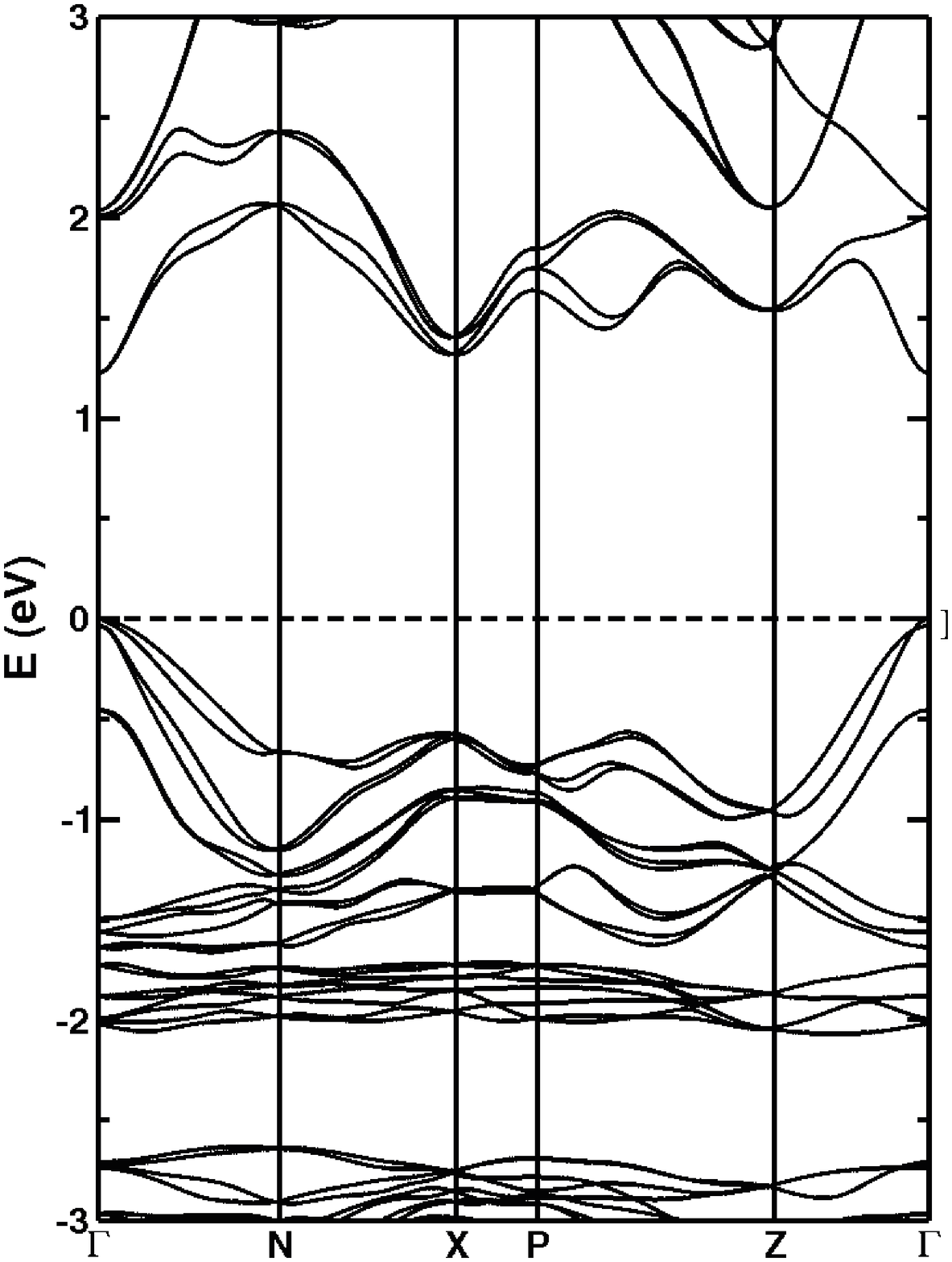}}\\
\subfigure[]{\includegraphics[width=65mm,height=65mm]{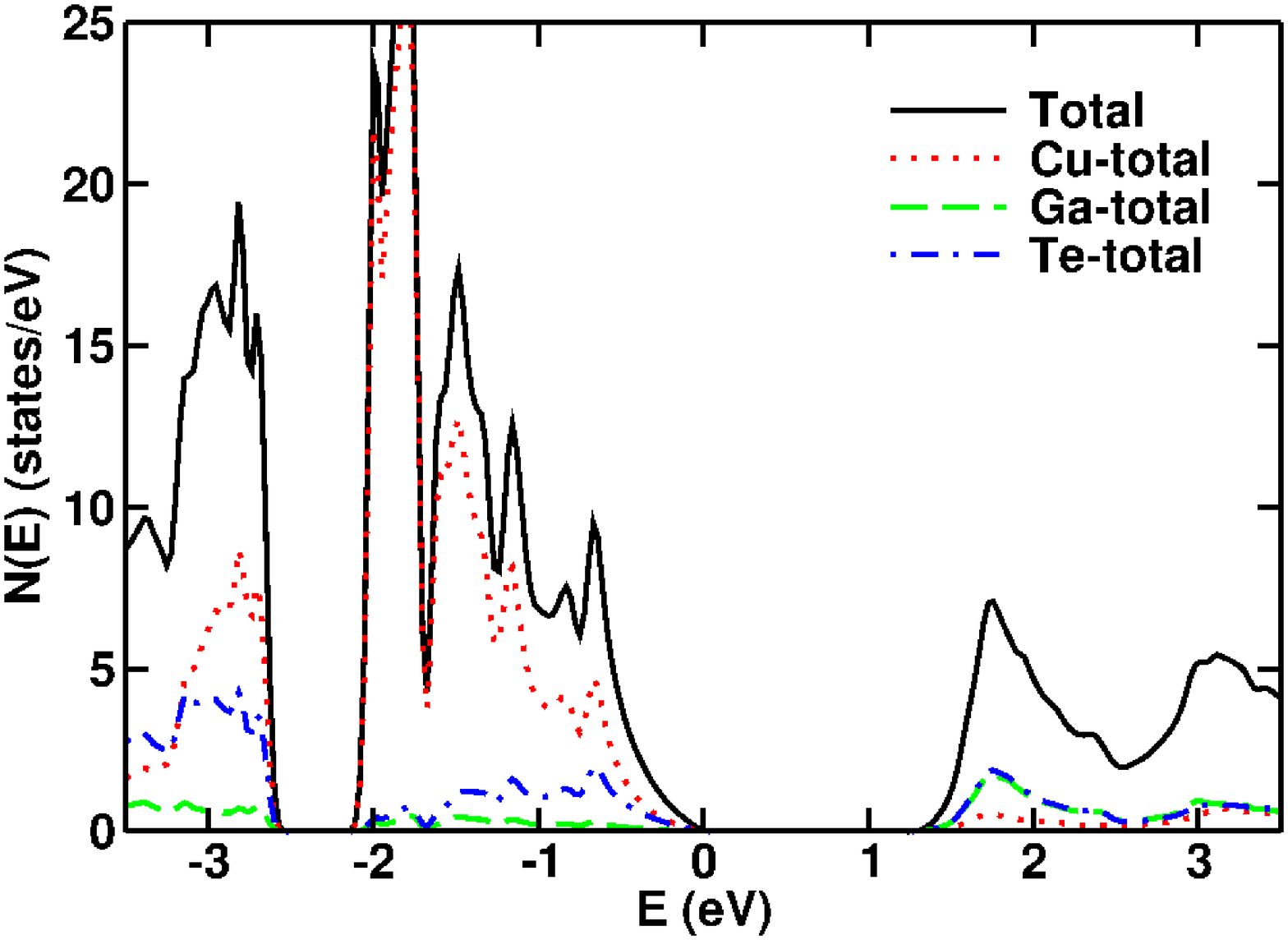}}
\caption{(Color online) Calulated (a) Band structure and (b) Density of states of CuGaTe$_2$. The energy zero corresponds to the valence band maximum.}
\end{center}
\end{figure}

\newpage
.

\begin{figure}
\begin{center}
\subfigure[]{\includegraphics[width=65mm,height=80mm]{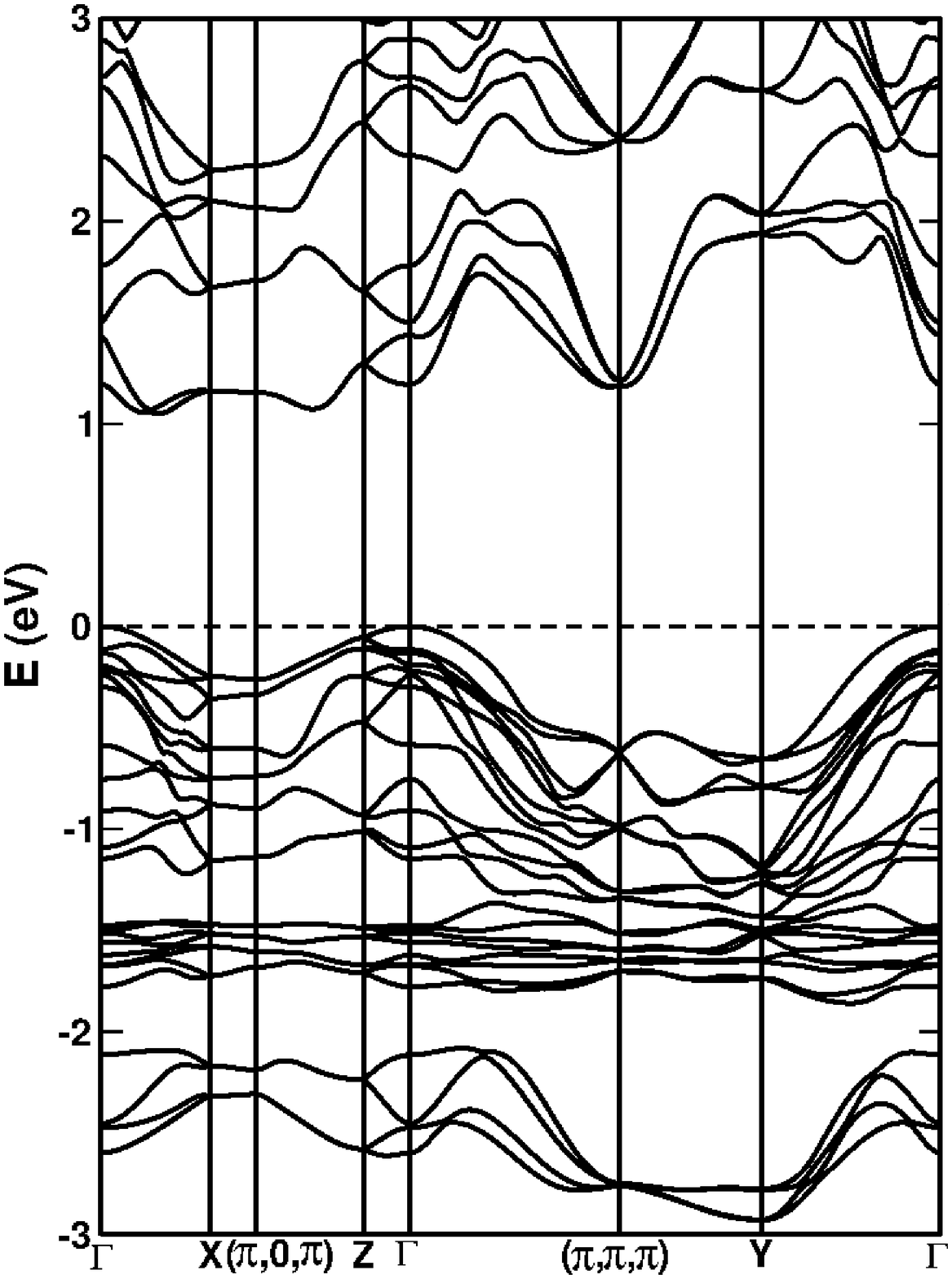}}\\
\subfigure[]{\includegraphics[width=65mm,height=65mm]{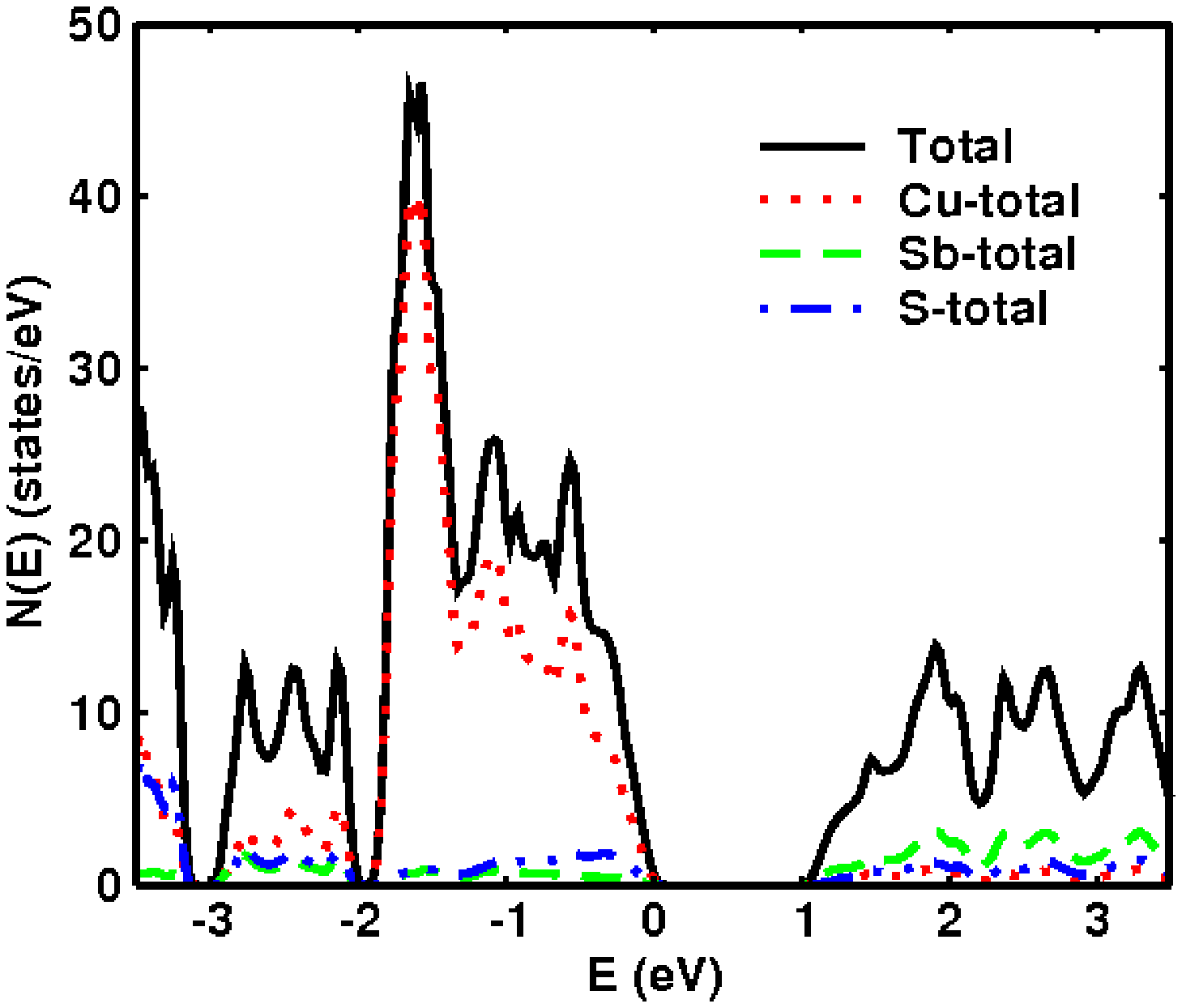}}
\caption{(Color online) Calulated (a) Band structure and (b) Density of states of CuSbS$_2$. The energy zero corresponds to the valence band maximum.}
\end{center}
\end{figure}

\begin{figure}
\begin{center}
\subfigure[]{\includegraphics[width=70mm,height=70mm]{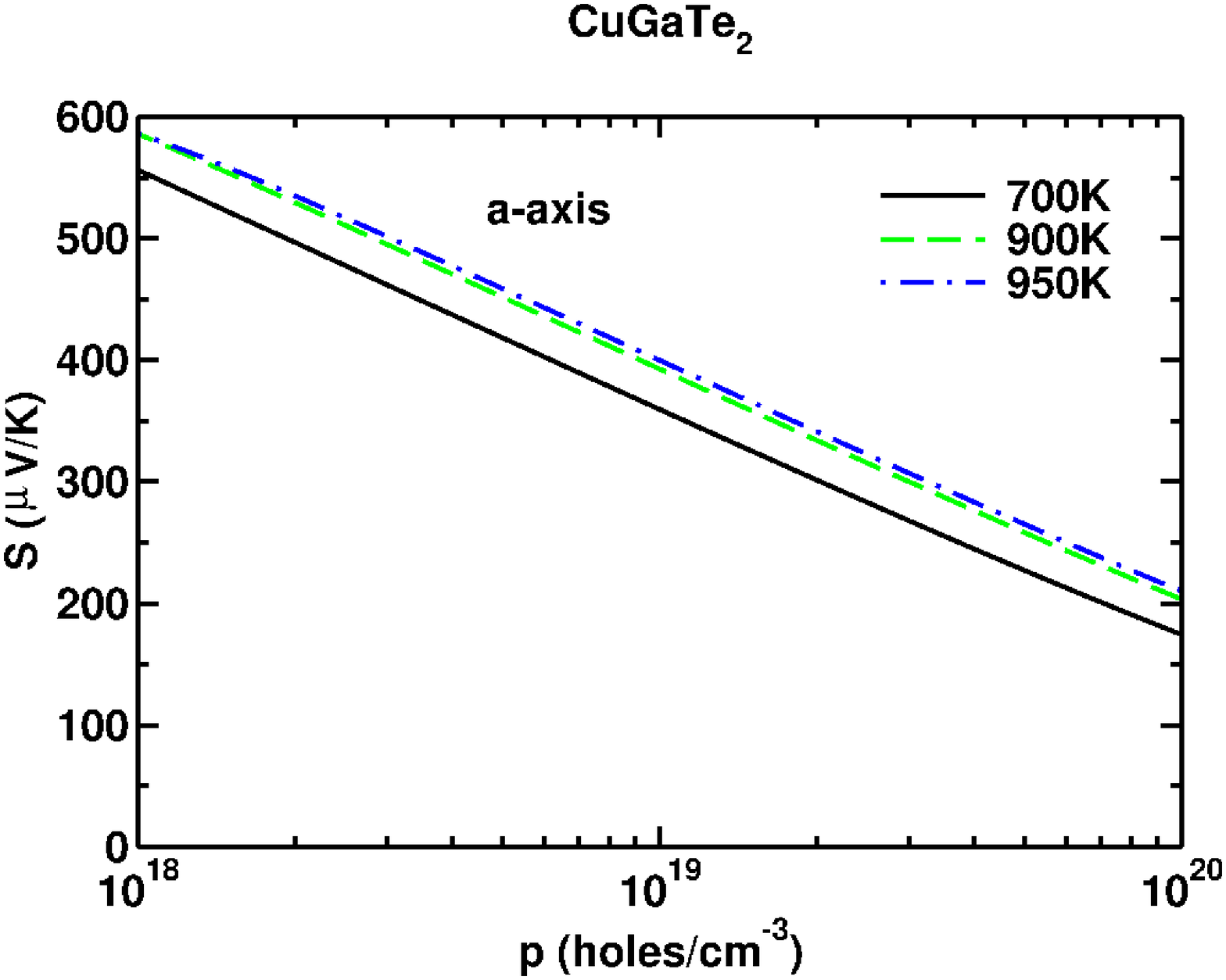}}
\subfigure[]{\includegraphics[width=70mm,height=70mm]{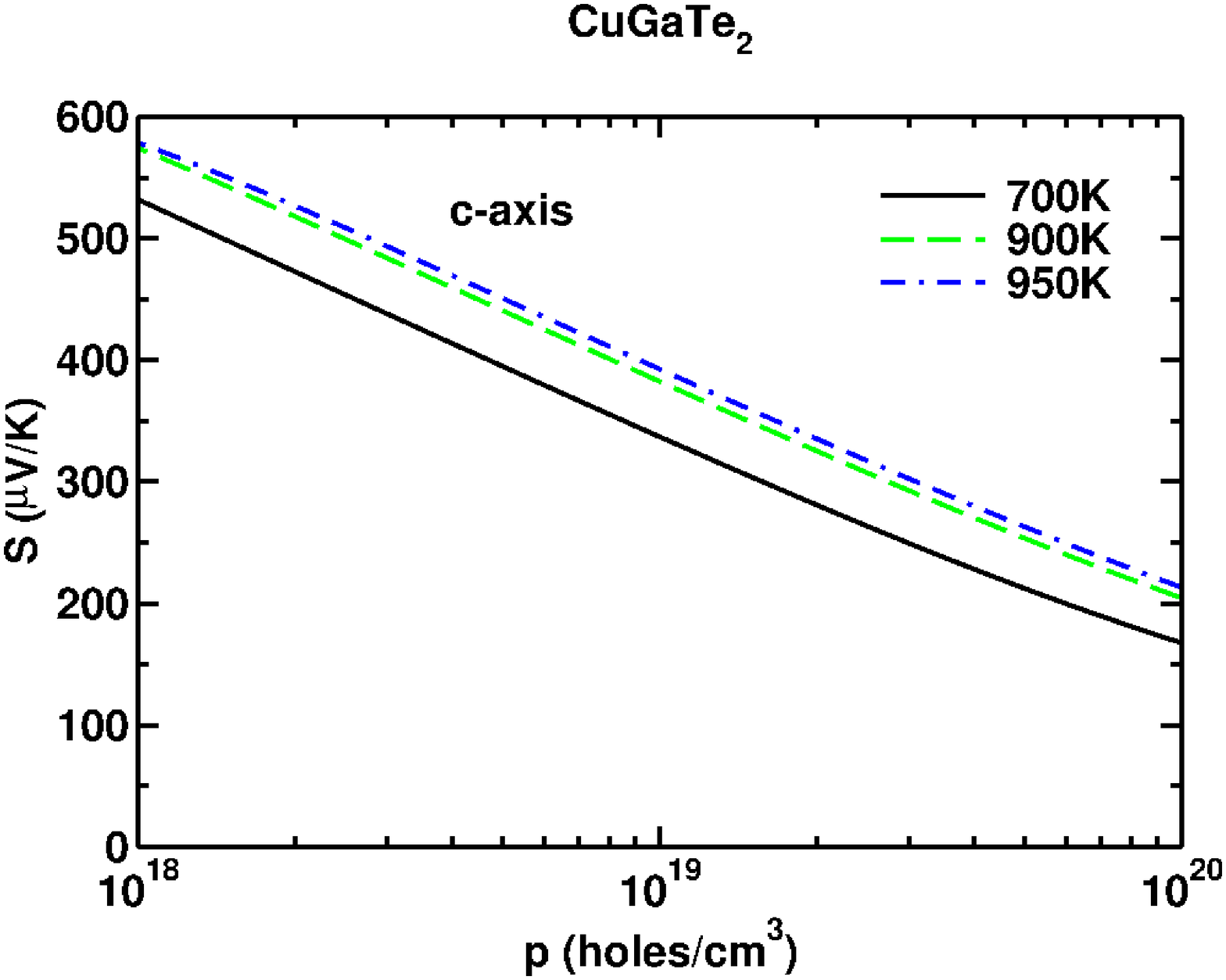}}
\caption{(Color online) Thermopower variation with hole concentration along the (a) a-axis and (b) c-axis at different temperatures for CuGaTe$_2$.}
\end{center}
\end{figure}

\begin{figure}
\begin{center}
\includegraphics[width=70mm,height=70mm]{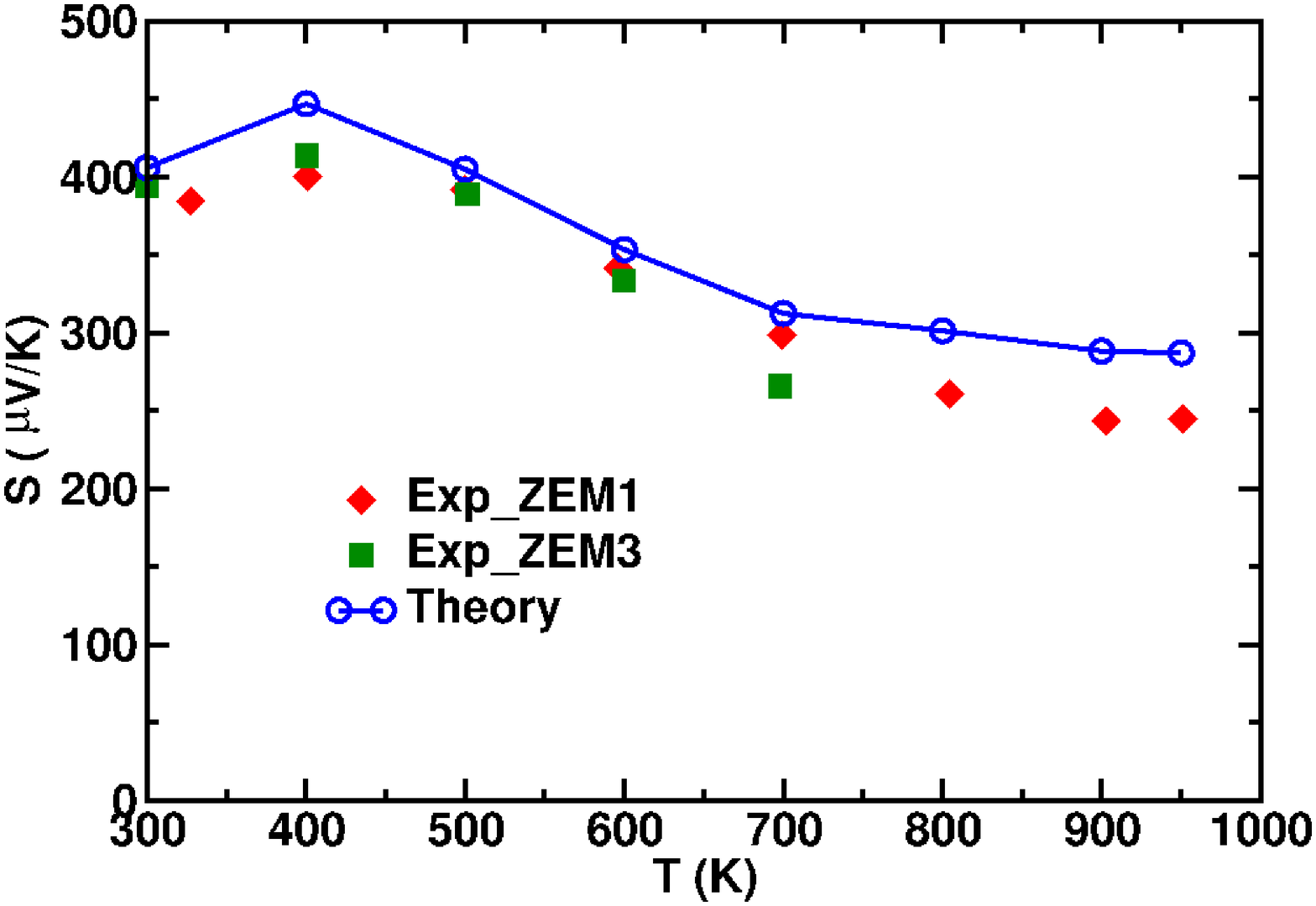}
\caption{(Color online) The variation of the calculated thermopower as a function of temperature for CuGaTe$_2$ (see text for explanation) 
compared with available experimental data (Ref. \onlinecite{ADV. MATER}) for two different samples. 
}
\end{center}
\end{figure}

\begin{figure}
\begin{center}
\includegraphics[width=70mm,height=70mm]{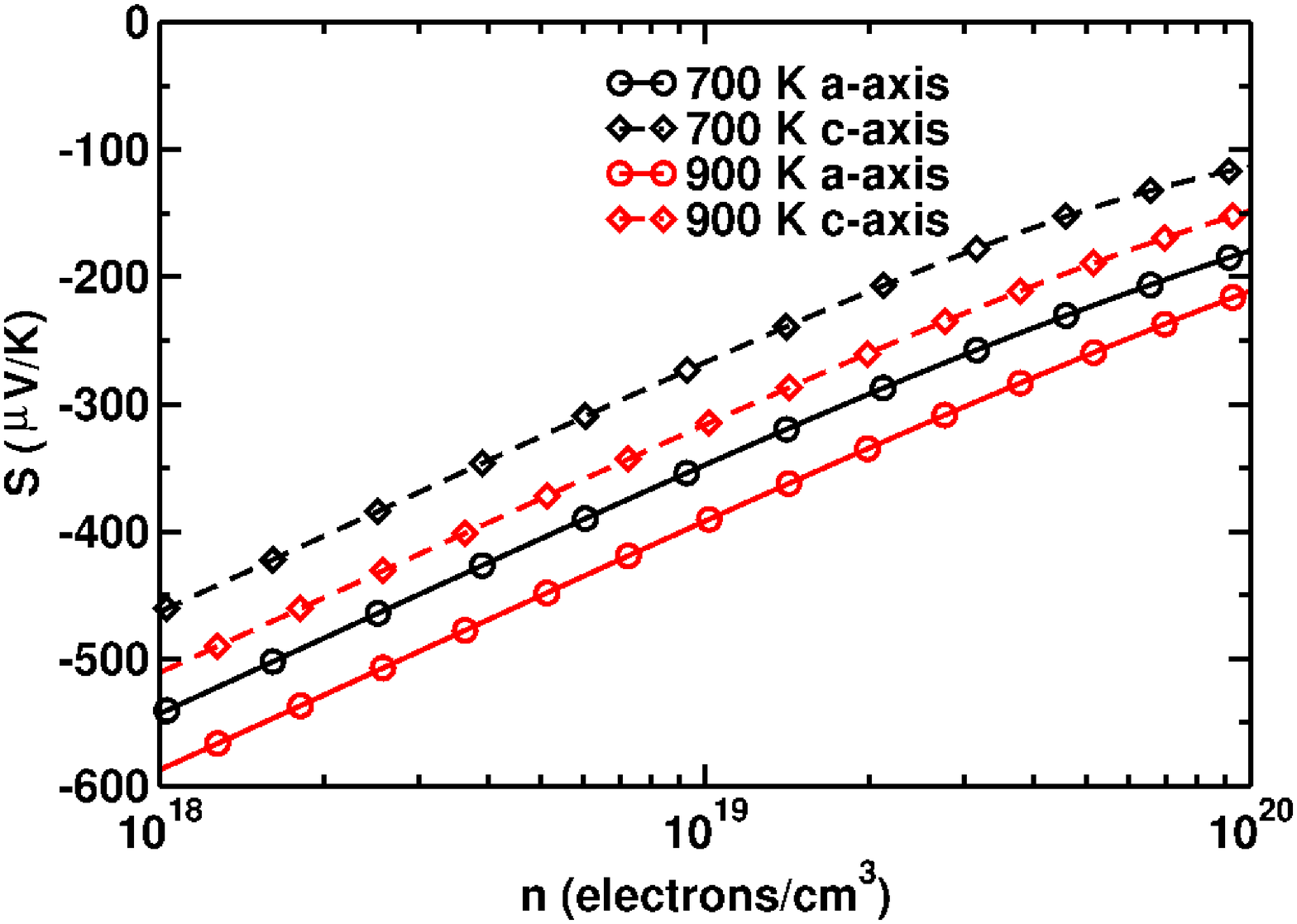}
\caption{(Color online) Thermopower variation with electron concentration along the $a$- and $c$-axes at different temperatures for CuGaTe$_2$.}
\end{center}
\end{figure}

\begin{figure}
\begin{center}
\subfigure[]{\includegraphics[width=70mm,height=70mm]{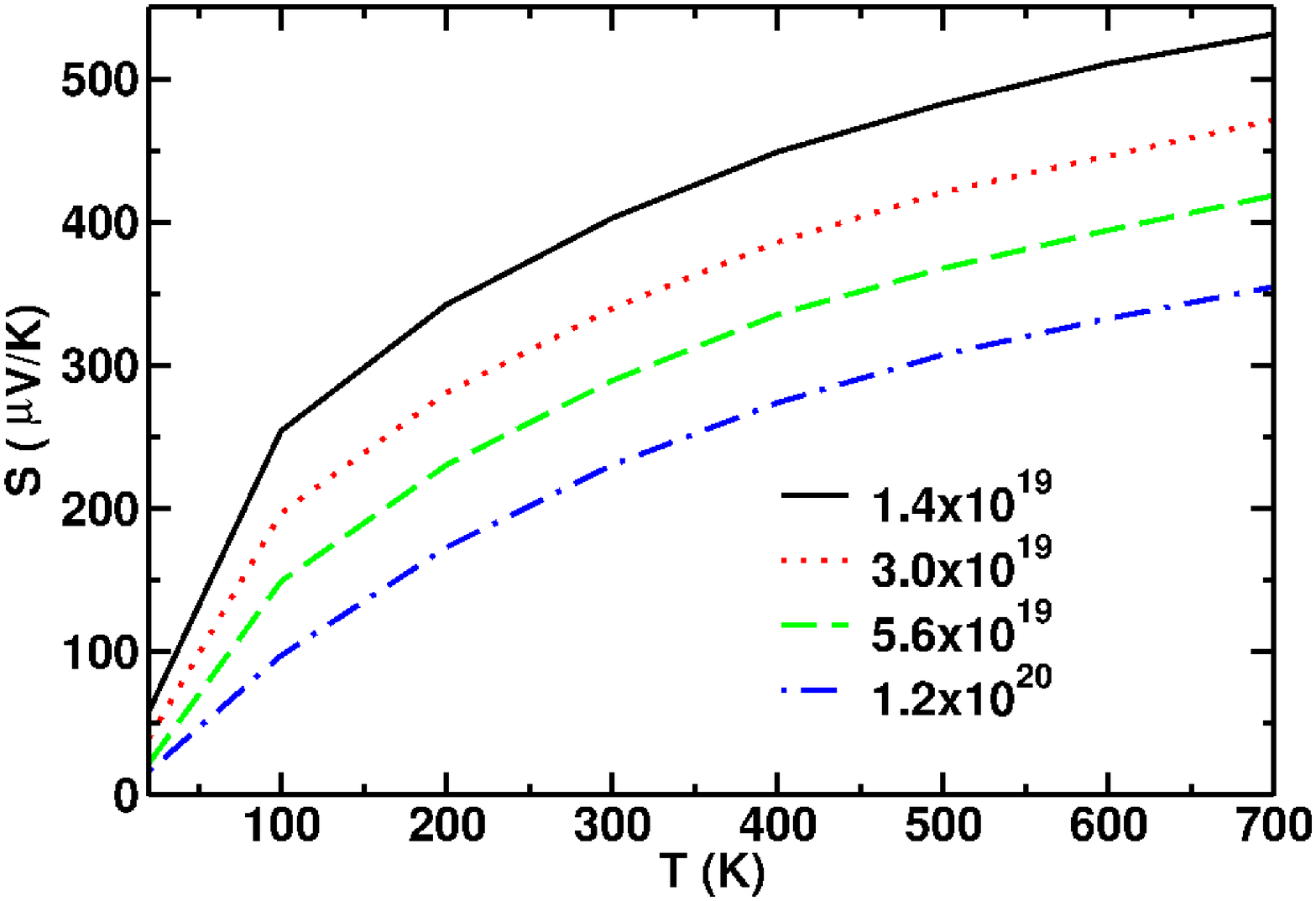}}\\
\subfigure[]{\includegraphics[width=70mm,height=70mm]{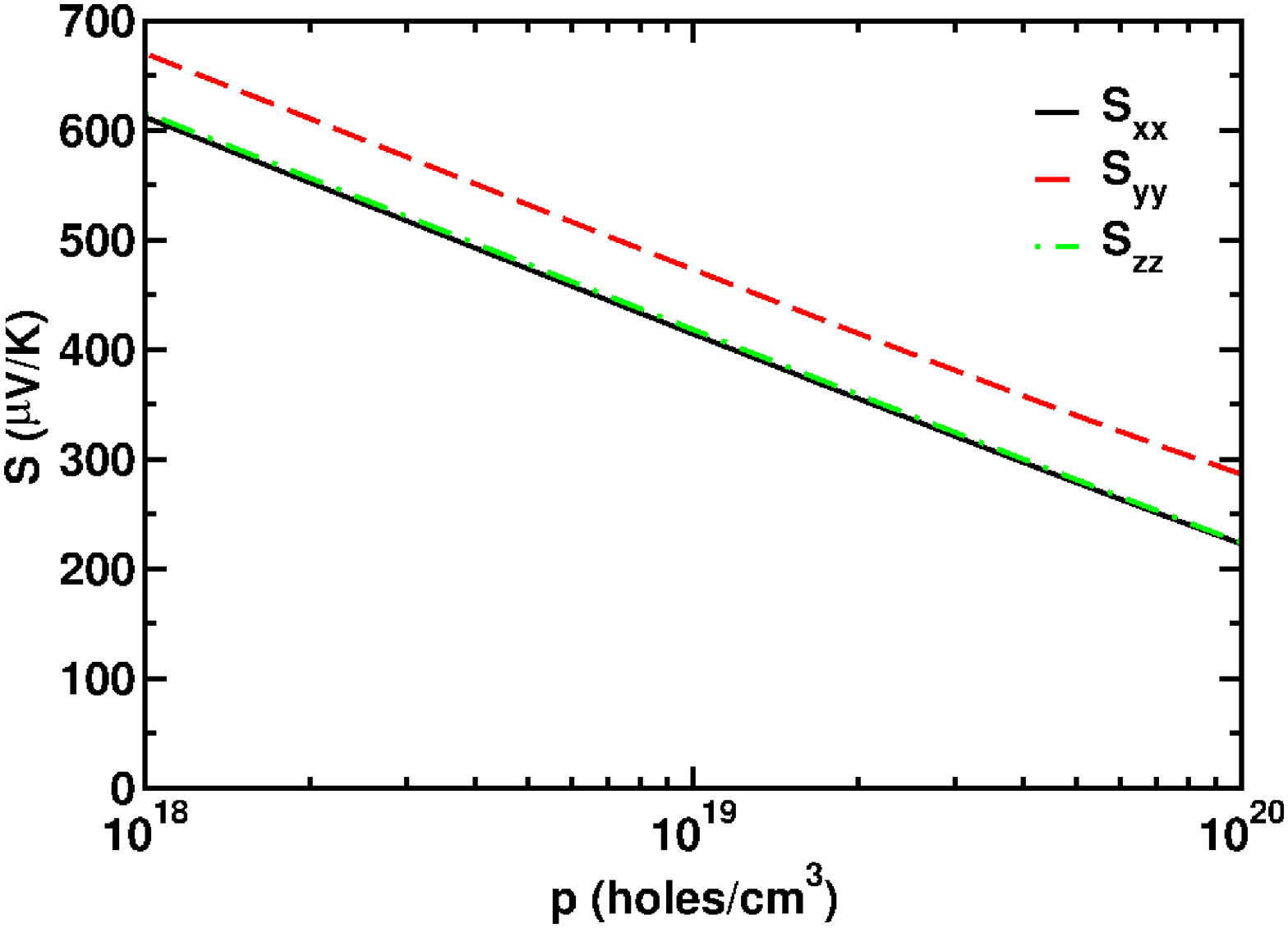}}
\subfigure[]{\includegraphics[width=70mm,height=70mm]{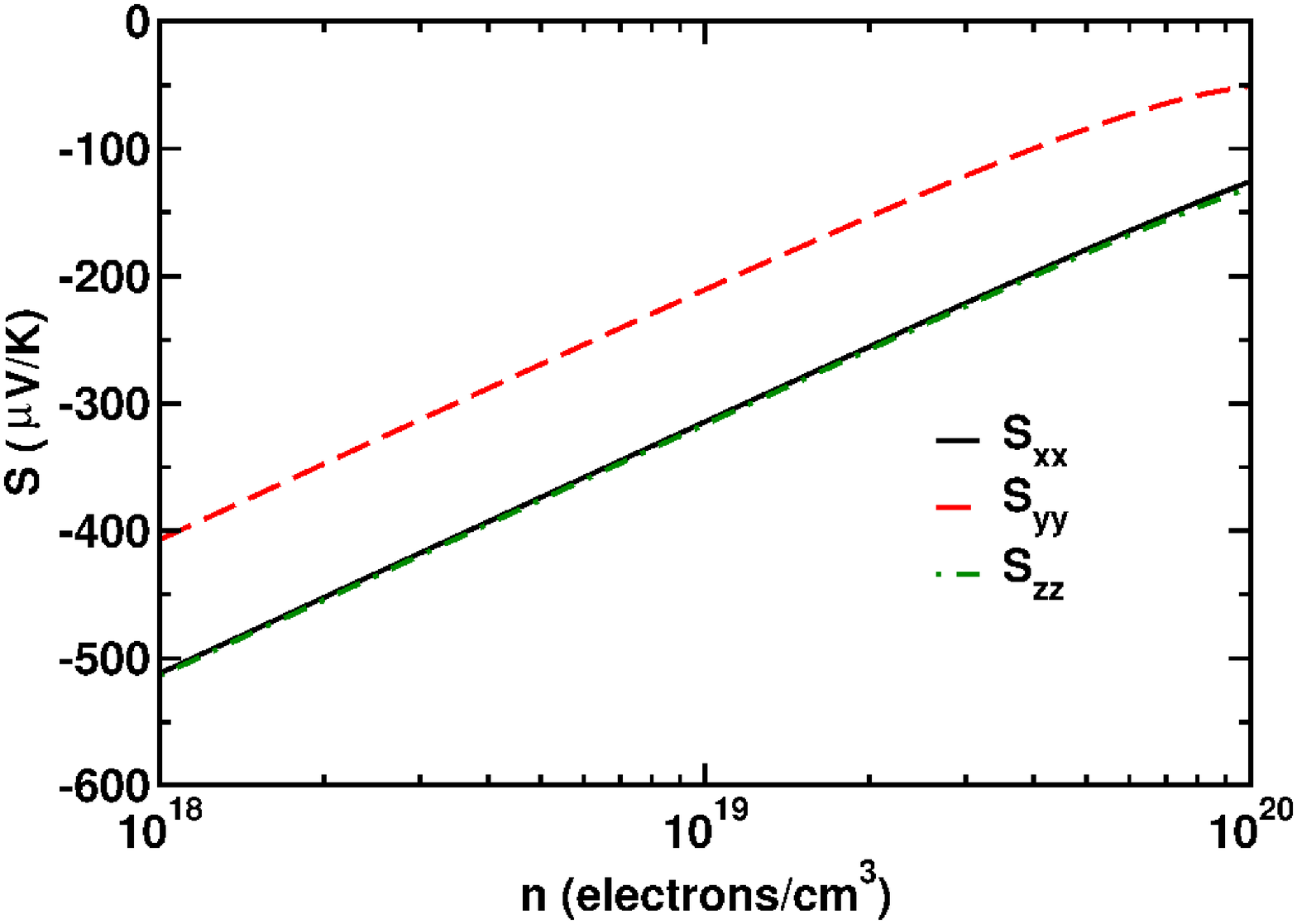}}
\caption{(Color online) (a) Thermopower variation($S_{xx}$) with temperature at different hole concentrations for CuSbS$_2$. 
Comparison of the thermopower variation with (b) hole concentration and (c) electron concentration for different directions, at temperature $T=300$ K.}
\end{center}
\end{figure}

\begin{figure}
\begin{center}
\includegraphics[width=70mm,height=70mm]{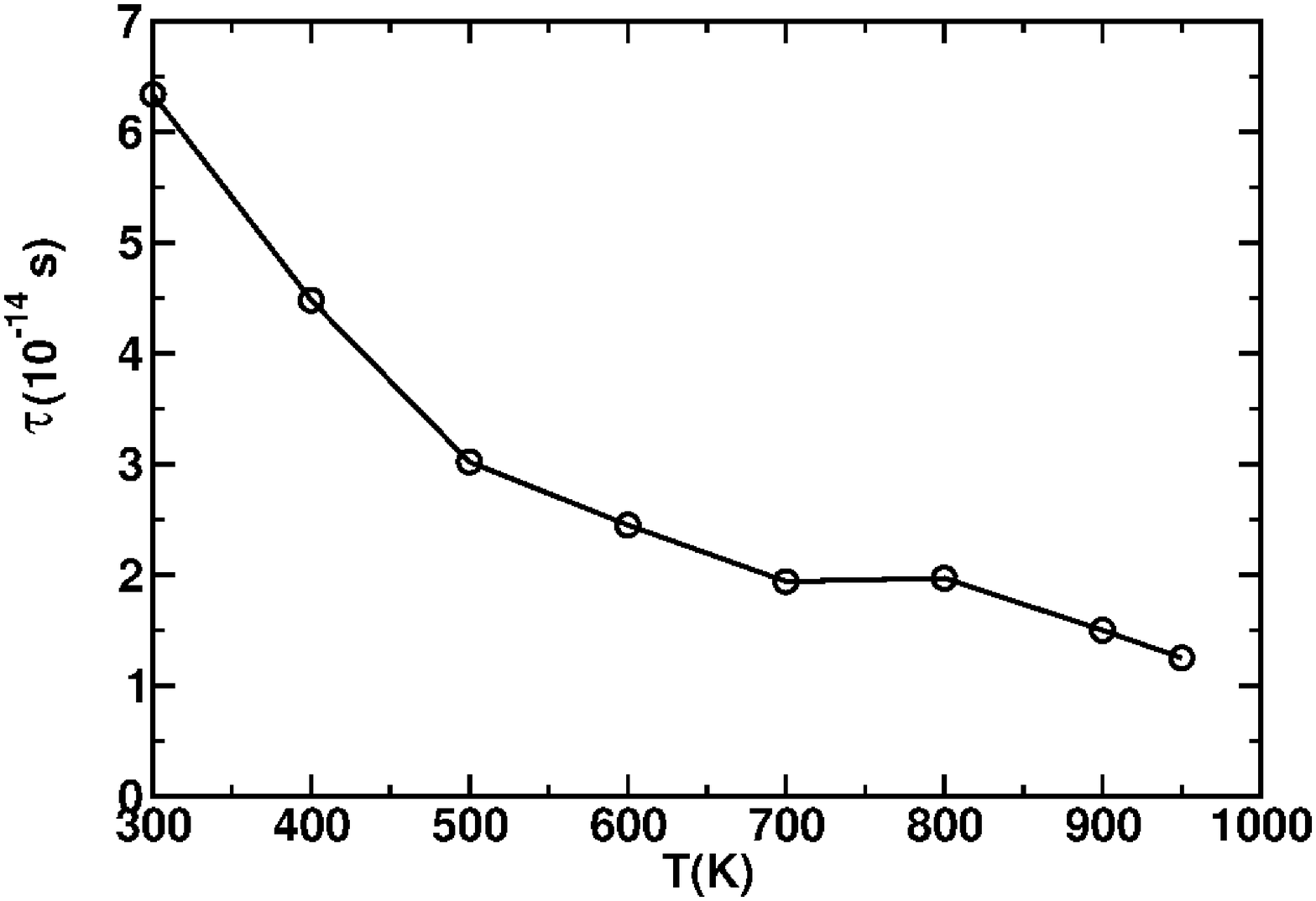}
\caption{ Temperature variation of the relaxation time $\tau$ of CuGaTe$_2$, estimated as explained in text.}
\end{center}
\end{figure}

\begin{figure}
\begin{center}
\subfigure[]{\includegraphics[width=70mm,height=70mm]{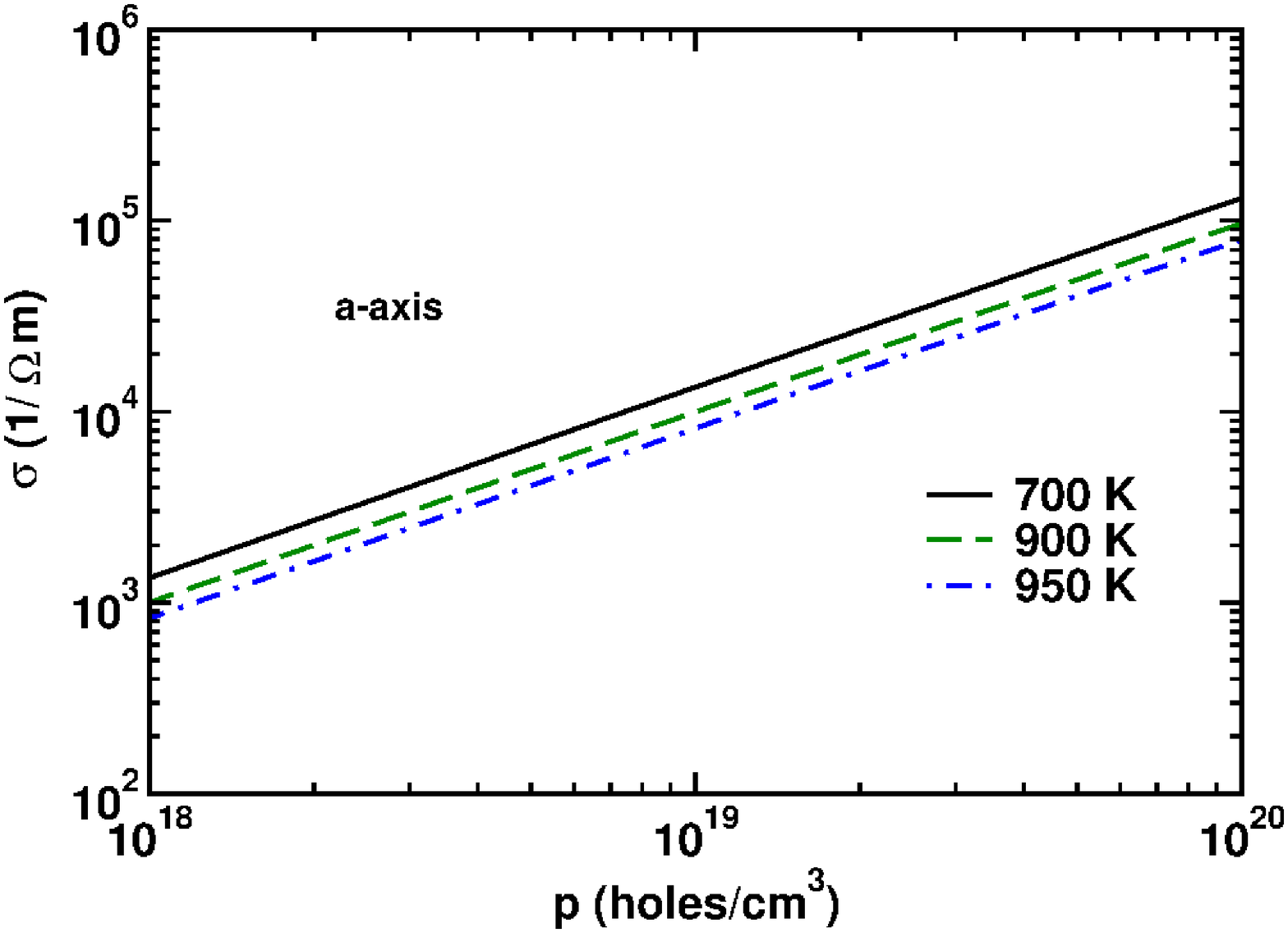}}
\subfigure[]{\includegraphics[width=70mm,height=70mm]{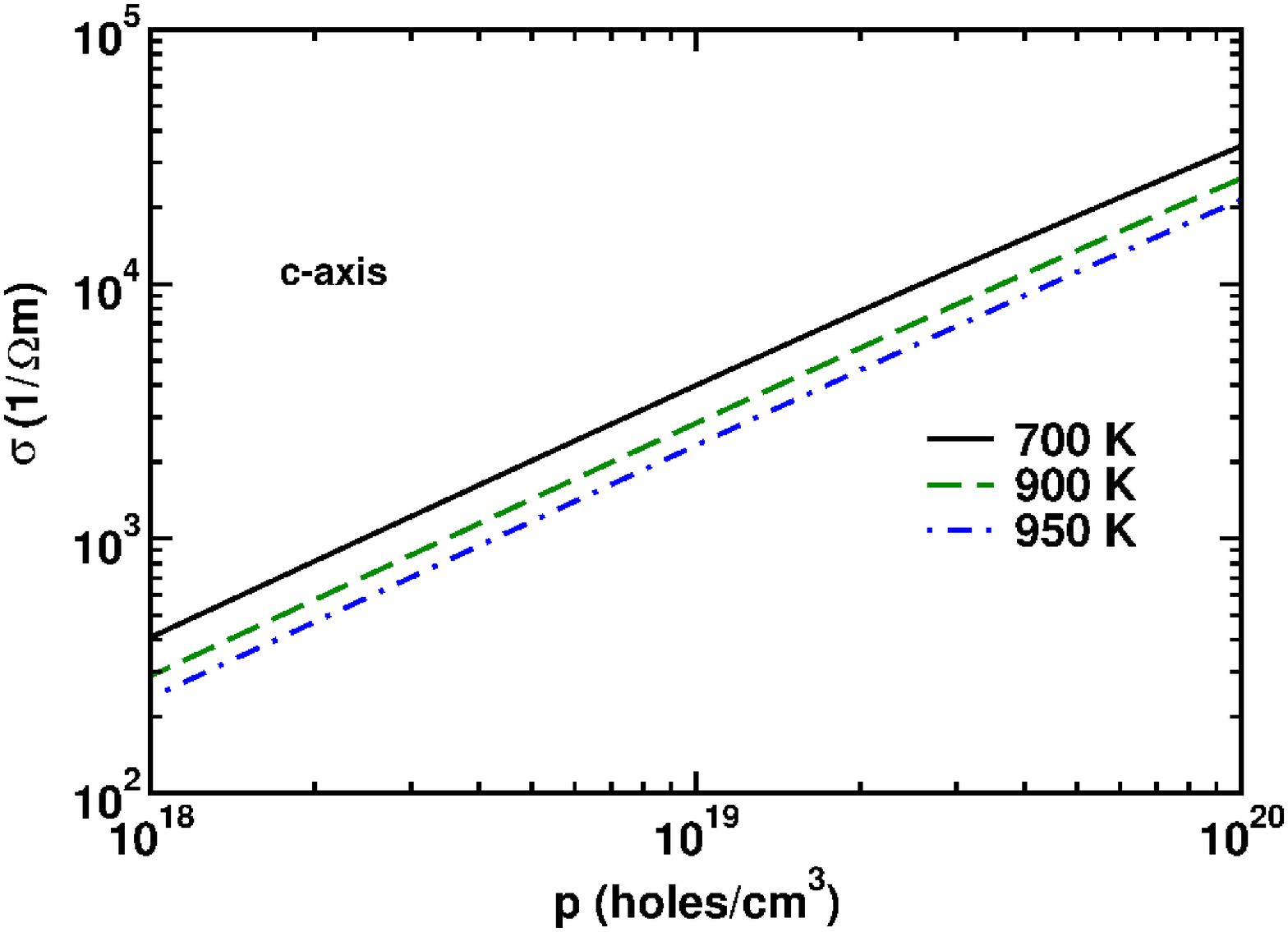}}
\caption{(Color online) Electrical conductivity variation with hole concentration along the (a) a-axis and (b) c-axis at different temperatures for CuGaTe$_2$.}
\end{center}
\end{figure}

\begin{figure}
\begin{center}
\includegraphics[width=70mm,height=70mm]{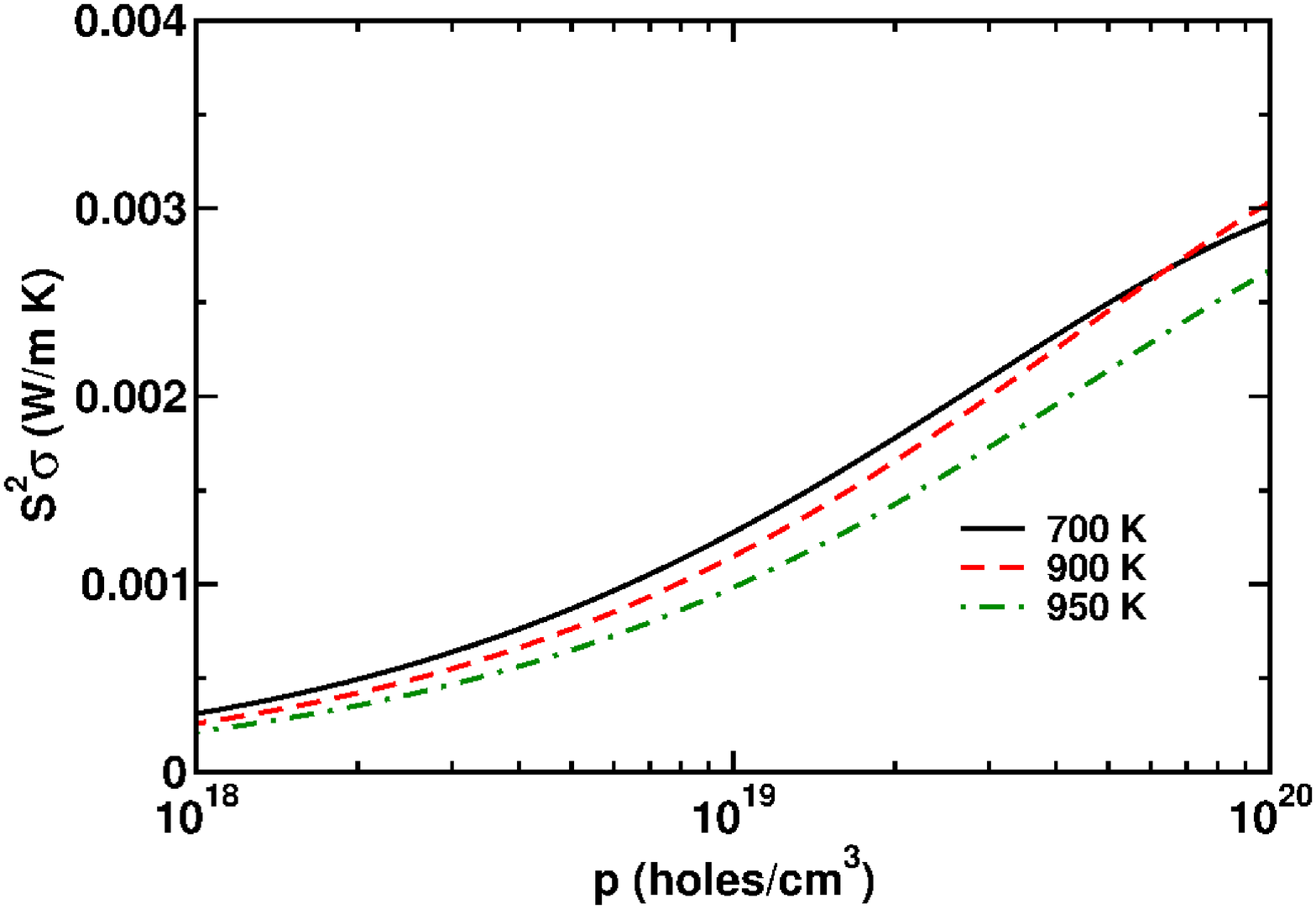}
\caption{(Color online) Variation of the power factor with hole concentration for CuGaTe$_2$.}
\end{center}
\end{figure}

\begin{figure}
\begin{center}
\subfigure[]{\includegraphics[width=70mm,height=70mm]{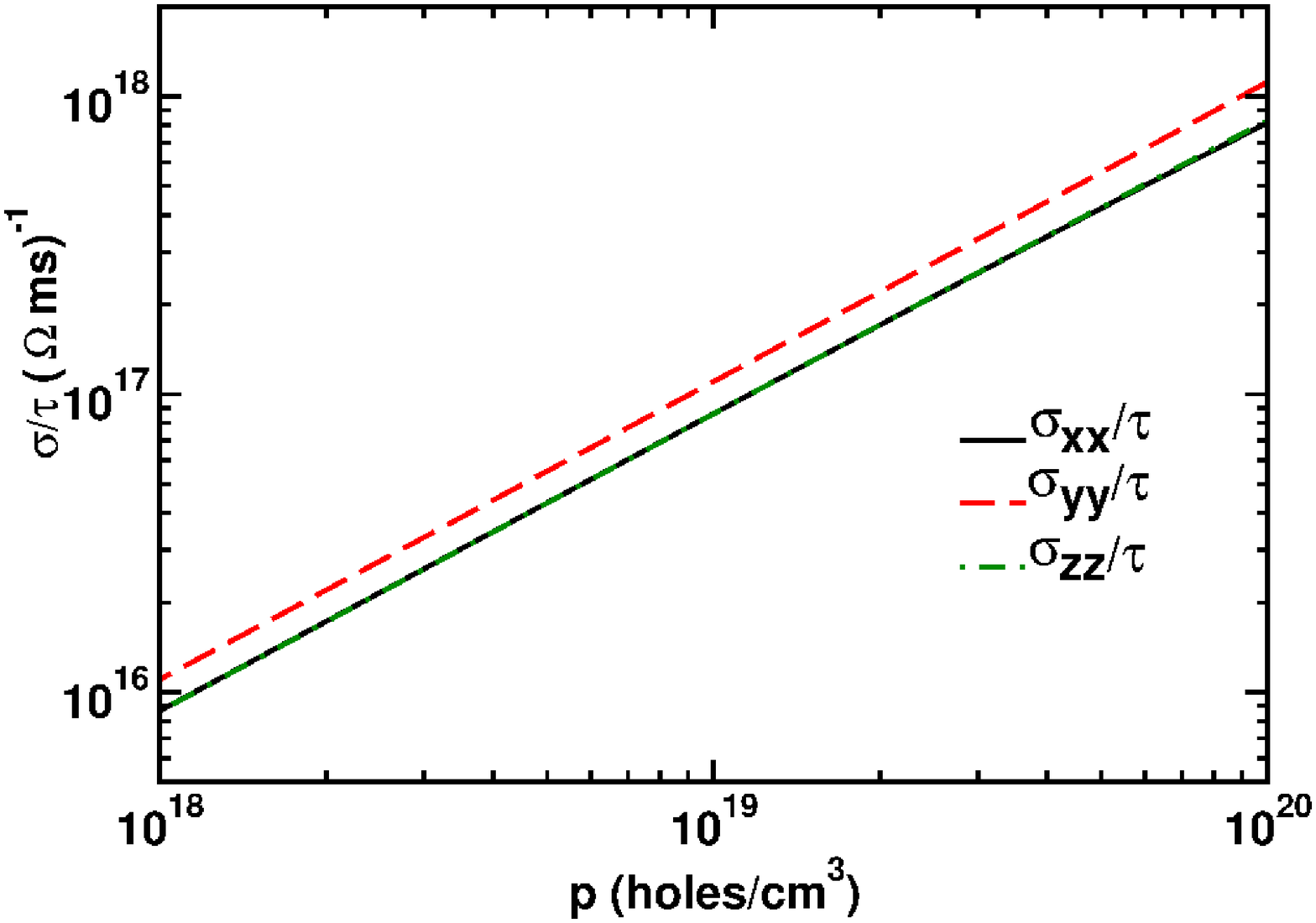}}
\subfigure[]{\includegraphics[width=70mm,height=70mm]{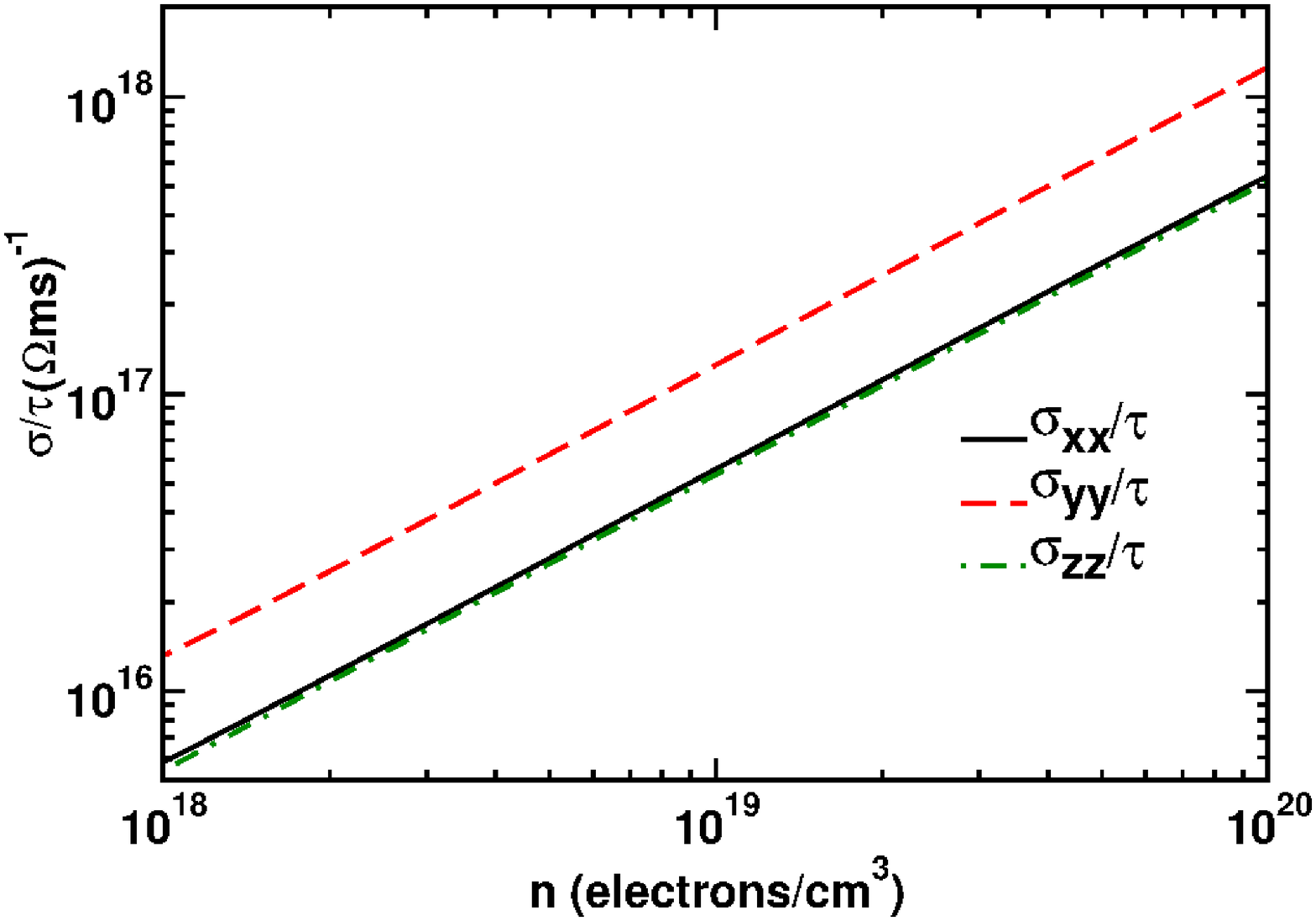}}
\caption{(Color online) Variation of the ratio $(\sigma$/$\tau)$ with (a) hole concentration and (b) electron concetration along the three lattice directions for CuSbS$_2$ at $T=300$ K.}
\end{center}
\end{figure}

\begin{figure}
\begin{center}
\includegraphics[width=70mm,height=70mm]{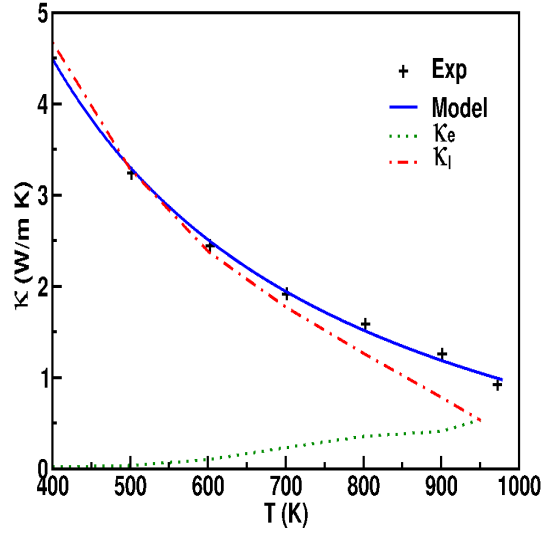}
\caption{(Color online) The experimental\cite{ADV. MATER} thermal conductivity of CuGaTe$_2$ (crosses) compared with the fitting expression (see text - full line, blue). The electronic (dashed) and lattice contributions (dash-dotted) are shown separately.}
\end{center}
\end{figure}

\begin{figure}
\begin{center}
\subfigure[]{\includegraphics[width=70mm,height=70mm]{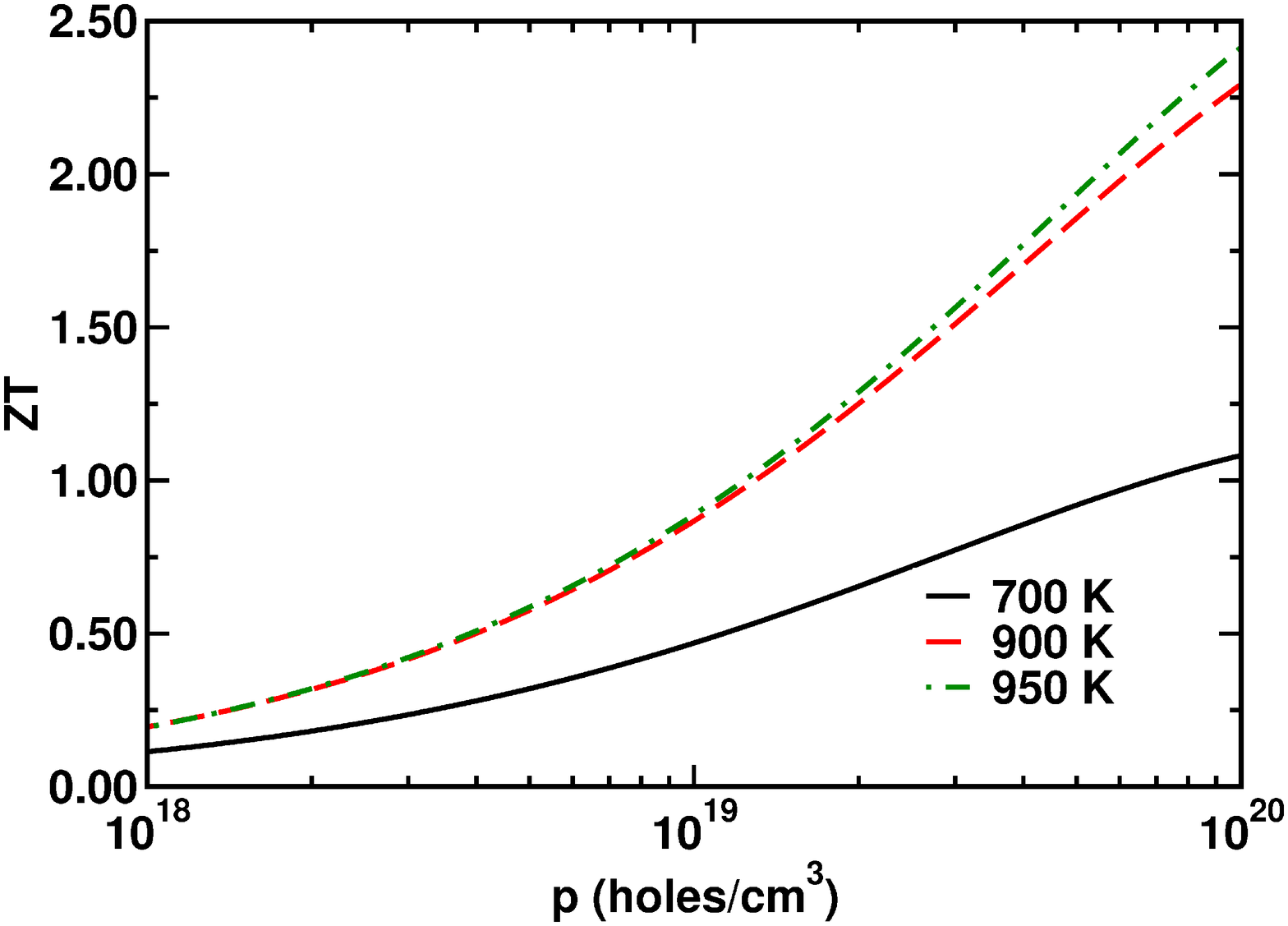}}
\subfigure[]{\includegraphics[width=70mm,height=70mm]{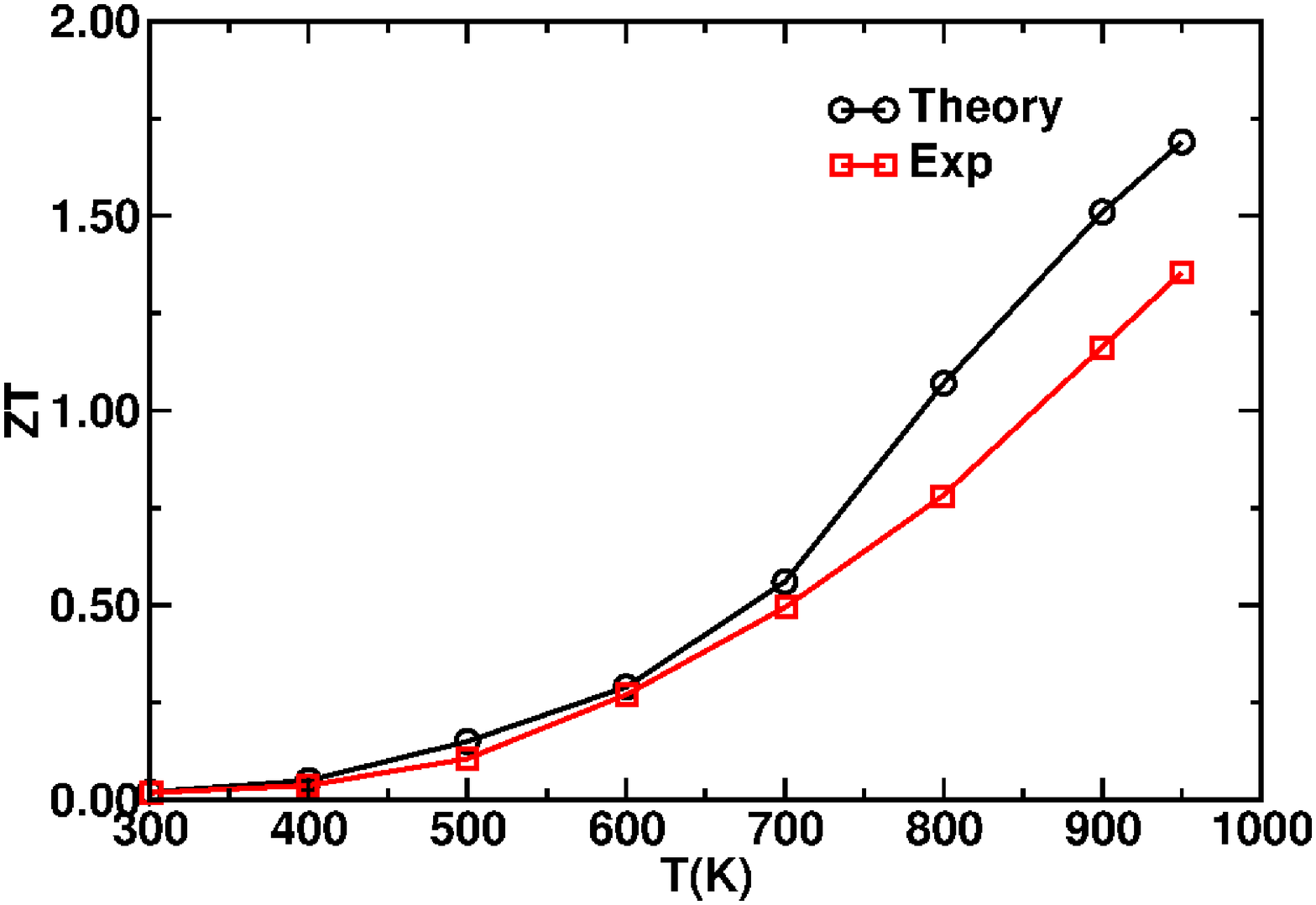}}
\caption{(Color online) Calculated figure of merit of CuGaTe$_2$ as function of (a) the hole concentration for three different temperatures
and (b) temperature for the experimental carrier density and compared to experiment.\cite{ADV. MATER}
}
\end{center}
\end{figure}

\end{document}